\documentclass[fleqn,reprint,groupedaddress,amsmath,amssymb,aps,prx,floatfix,longbibliography]{revtex4-1}
\usepackage{graphicx}
\usepackage{amsmath}
\usepackage{multirow} 
\usepackage{physics} 
\graphicspath{{Figures/}}

\begin{document}

\title{Microscopic control and detection of ultracold strontium in optical-tweezer arrays}

\author{M. A. Norcia}
\author{A. W. Young}
\author{A. M. Kaufman}
\email[E-mail: ]{adam.kaufman@colorado.edu}
\address{JILA, University of Colorado and National Institute of Standards and Technology, and
Department of Physics, University of Colorado, Boulder, Colorado 80309, USA}

\begin{abstract} 

We demonstrate a set of tools for microscopic control of neutral strontium atoms.  We report single-atom loading into an array of sub-wavelength scale optical tweezers, light-shift free control of a narrow-linewidth optical transition, three-dimensional ground-state cooling, and high-fidelity nondestructive imaging of single atoms on sub-wavelength spatial scales. Extending the microscopic control currently achievable in single-valence-electron atoms to species with more complex internal structure, like strontium, unlocks a wealth of opportunities in quantum information science, including tweezer-based metrology, new quantum computing architectures, and new paths to low-entropy many-body physics.

\end{abstract}

\date{\today}

\maketitle

Quantum systems comprised of neutral atoms have been a successful vehicle for studies in information processing, metrology, and simulation~\cite{Weiss2017,Bloch2008,YeReview}.  This success is, in part, due to the capacity of neutral atoms to be initialized in large, isolated quantum states~\cite{Bloch2008}, as well as their compatibility with microscopic observables~\cite{Gross2017}. With high-resolution optics, single atoms bound in micron-scale arrays may be resolved through fluorescence detection~\cite{Schlosser2001, Bakr2009}, enabling parallel qubit readout and new perspectives on many-body quantum systems~\cite{Cheneau2012, Xia2015, Kaufman2016, Mazurenko2017}. In an emerging frontier, platforms like quantum gas microscopes and optical tweezers harness microscopy techniques not just for detection, but to create quantum states at vanishing entropy through microscopic control. The generation of ultralow-entropy systems through atom-by-atom assembly, atomic cookie-cutting, and entropy redistribution~\cite{Weitenberg2011, Preiss2013, Kaufman2014,  Barredo2016, Endres2016, Mazurenko2017,Kumar2018} has led to ground-breaking studies in atomic quantum optics, many-body entanglement, and simulations of quantum magnetism~\cite{Kaufman2014,Islam2015, Bernien2017,Lienhard2018}. 

\begin{figure}[!htb]
\includegraphics[width=\linewidth, ]{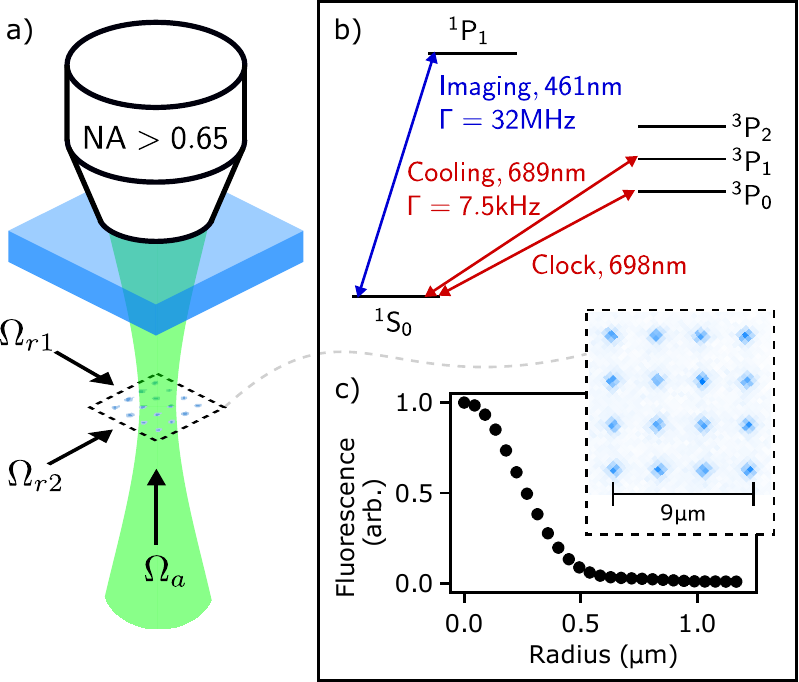}
\caption{
(a) Apparatus for microscopic control of arrays of strontium atoms.  We use a high numerical aperture (NA $>$ 0.65) objective to focus 515~nm light into arrays of optical tweezers in which we trap and manipulate individual strontium atoms.  (b) Strontium atoms have transitions with a wide range of linewidths, which are useful for imaging ($^1$S$_0$  to $^1$P$_1$), cooling ($^1$S$_0$  to $^3$P$_1$) and for optical frequency metrology ($^1$S$_0$  to $^3$P$_0$). In this work we probe the cooling transition in the axial direction with beam $\Omega_a$ and in two near-orthogonal radial directions with beams $\Omega_{r1,r2}$. (c) Point-spread function (PSF) from single atoms in our optical tweezers, corresponding to an effective Gaussian waist of 0.44(2)~$\mu$m (black points). In (a, c) averaged atom array images are displayed in the dashed boxes. In this work, we use 3$\times$3 or 4$\times$4 arrays with a lattice constant of $\mathrm{\sim3\ \mu m}$ to ensure that our atoms behave independently, however, our optical system is compatible with substantially larger arrays with tighter spacings~(see supplement \cite{Supplement}).}
\label{fig:setup}
\end{figure}

So far, much of the microscopy work with neutral atoms has focused on alkali species, which have a single valence electron~\cite{BlochQGM, GreinerQGM, Li2012, Kaufman2012, Omran2015, Greif2016, Cheuk2016, Endres2016, Barredo2016, Kumar2018}. The increased complexity of species like the two-valence-electron alkaline-earth atoms and molecules brings new opportunities --- new quantum computation architectures~\cite{Daley2008, Weitenberg2011a, pagano2018fast}, new types of spin models~\cite{Feig2010a, Hermele2009,Ozawa2018},  premier time-keeping~\cite{campbell2017fermi,schioppo2017ultrastable,ushijima2015cryogenic, poli2009simplified} --- which have spurred the development of techniques for the microscopic detection and control of such particles~\cite{yamamoto2016ytterbium, Miranda2017, Chou2017, liu2018building}. Here, through the combined use of optical tweezer arrays, high-resolution imaging, and narrow-line spectroscopy, we show single-particle preparation and detection of the alkaline-earth atom strontium (Fig.~\ref{fig:setup}a,b).  

The ability to image strontium on sub-wavelength scales (Fig.~\ref{fig:setup}c) and in shallow potentials, combined with its nuclear and electronic degrees of freedom, enables exciting new possibilities. For example, the imaging techniques presented here may enable microscopy of large-spin SU(10) magnetism and heavy fermion Kondo physics~\cite{Stewart1984, Feig2010, Feig2010a, Hermele2009, Balents2010, Hofrichter2016, Ozawa2018}. Meanwhile, the tweezer-trapping we show could pave the way for new computing architectures based on spin-orbital exchange gates~\cite{Daley2008, Weitenberg2011a, pagano2018fast}, as well as explorations of Rydberg many-body physics in a two-electron atom~\cite{Bannasch2013, Labuhn2016, Bernien2017, Pohl2014}.

In this work, we describe how strontium's optical transitions can deliver high-fidelity cooling in a sub-wavelength scale optical tweezer. Using $^{88}\text{Sr}$, we achieve a three-dimensional ground-state occupancy of $91^{+9}_{-25}$~\% in $0.48(2)~\mu$m waist tweezers. Such high-fidelity laser cooling has been demonstrated in optically-trapped alkalis~\cite{Li2012, Kaufman2012, cheuk2015quantum, Yu2018} using two-photon Raman transitions and optical pumping. However, this approach requires careful consideration of beam geometries, polarizations, and excited state potentials. By contrast, the $\mathrm{{^1S_0} \leftrightarrow {^3P_1}}$ intercombination line of strontium represents a nearly ideal two-level system, with a 7.5 kHz linewidth amenable to fast but high-fidelity cooling given our degree of confinement~\cite{Blatt2009}. Such narrow-linewidth optical transitions demand careful control of light shifts that can spoil the spectral resolution required for cooling. To this end, we characterize a technique to significantly suppress these shifts in deep tweezers compatible with ground-state cooling~\cite{ido2003recoil,yamamoto2016ytterbium}. On its own, this tweezer-based ground state cooling may be used to reduce motional dephasing effects present in Rydberg-based two-qubit gates~\cite{Saffman2010, Levine2018}, and can aid new directions in metrology, such as tweezer-based optical atomic clocks.

When combined with a wavelength-scale optical lattice, the full set of capabilities demonstrated here may enable a new path to microsopic preparation and control of low-entropy many-body systems. In this approach, tweezers may be used to rapidly assemble arbitrary initial conditions with vanishing entropy through lossless imaging, rearrangement, and ground-state laser cooling \cite{Kaufman2014, Endres2016, Barredo2016}. Sub-wavelength optical tweezers will allow state-preserving transfer into an optical lattice potential, which provides a highly coherent Hubbard-type system typically accessed through evaporative cooling. This new approach would prove advantageous for studies of many-body phenomena at low energy scales --- such as magnetism --- for which tailored state preparation can be useful~\cite{Lubasch2011, Chiu2018}, for scaling of protocols to measure and quantify entanglement~\cite{Islam2015, Kaufman2016, Brydges2018}, as well as for novel directions like atom-based studies of sampling problems~\cite {Harrow2017, Deutsch2017}.

\begin{figure*}[!htb]
  \centering
  \includegraphics[width=.9\textwidth]{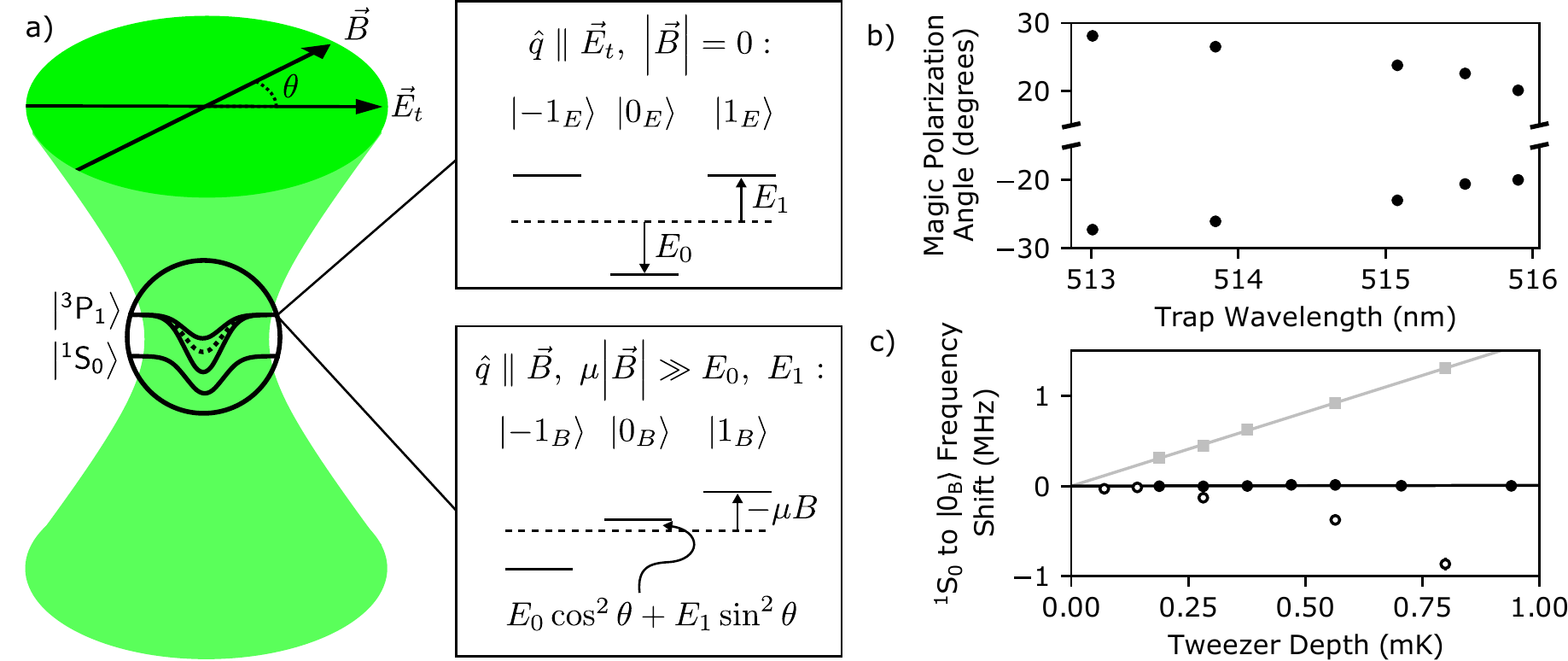}
  \caption{Magic polarization angle.  (a) By setting the angle $\theta$ between an applied magnetic field ($\vec{B}$) and the polarization of the tweezer light ($\vec{E}_t$), we can achieve equal trap depths between the ground ($^1$S$_0$) and an optically excited ($^3$P$_1$) state.  In the absence of a magnetic field, the tweezer light causes energy shifts to the different spin projections of $^3$P$_1$. These shifts are of opposite sign relative to the shift of the ground state (dashed lines denote the shifted ground state offset by the unperturbed transition frequency). By applying an appropriately oriented magnetic field, we can create an eigenstate $\ket{0_B}$ for which these shifts cancel, leading to state-insensitive trapping.  We define the ``magic'' polarization angle $\theta$ for which the frequency of the $^1$S$_0$ to $\ket{0_B}$ transition is unperturbed by the tweezer depth (for sufficiently low depths relative to the field strength).  Full details are provided in the main text.  (b) Magic polarization angle as a function of tweezer wavelength, showing that a magic condition exists over a range of wavelengths.  (c) Shifts of the $^1$S$_0$ to $\ket{0_B}$ transition as a function of tweezer depth for a large (22 Gauss) field oriented at $\theta \simeq 0$ (gray squares), a large (24 Gauss) field oriented at the magic angle (black points), and a small (7 Gauss) field oriented at the magic angle (empty circles). All data in (c) is at a tweezer wavelength of 515.13~nm.  Grey, black lines are linear fits.
}
\label{fig:magic}
\end{figure*}

\section{Magic-field spectroscopy in optical tweezers}

To begin our experiments, we load strontium atoms from a magneto-optical trap formed by sawtooth-wave adiabatic passage cooling \cite{norcia2018narrow,muniz2018robust} on the 7.5~kHz linewidth $\mathrm{{^1S_0} \leftrightarrow {^3P_1}}$ transition into a two-dimensional array of tweezers.  Typically, we use either 3$\times$3 or 4$\times$4 arrays spaced by roughly 5~$\mu$m, so the atoms in individual tweezer spots behave independently from one-another.  Light-assisted collisions between atoms that occupy the same tweezer lead to pairwise loss, which results in either non-occupied or singly-occupied tweezers with roughly equal probability~\cite{Schlosser2001}.  

In order to harness the narrow linewidth $^3$P$_1$ excited state for effective cooling, it is critical that the tweezer potential experienced by the ground and excited states be equal, so that the frequency of the transition does not depend on the depth of the tweezer or the location of the atom within the tweezer.  In optical lattice clocks, a specific wavelength of trapping light (known as a ``magic" wavelength) is used to generate such equal shifts on an ultra-narrow ``clock" transition~\cite{Ye1734}.

Once the atoms are loaded, we create a similar magic trapping condition on the $\mathrm{{^1S_0} \leftrightarrow {^3P_1}}$ transition by applying a magnetic bias field at a specific (wavelength dependent) angle from the tweezer polarization. This concept has been previously used to demonstrate state-insensitive trapping in strontium \cite{ido2003recoil} and ytterbium \cite{yamamoto2016ytterbium} in optical lattice potentials. The potentials that we use in our tweezer system are necessarily deep in order to provide tight three-dimensional confinement, so care must be taken to maintain state insensitive trapping in the presence of these large shifts.  Here, we systematically characterize the application of these techniques to deep tweezers, which can lead to a departure from the perturbative regime previously considered. 

The $^3$P$_1$ state has a total spin of $J=1$, and thus three Zeeman sublevels (Fig.~\ref{fig:magic}a).  In the absence of a magnetic field or a circularly polarized component of the tweezer light, two of these sublevels (which we label $\ket{-1_E}$ and $\ket{1_E}$, expressed with respect to a quantization axis $\hat{q}$ oriented along the tweezer polarization $\vec{E}_t$) shift by the same amount, $E_1$, relative to the ground state $^1$S$_0$.  The third state $\ket{0_E}$ shifts by a different amount $E_0$ relative to $^1$S$_0$.  If $E_1$ and $E_0$ have opposite signs, we can use an appropriately oriented magnetic field to mix the shifts associated with the $\ket{0_E}$ state with those of the $\ket{-1_E}$ and $\ket{1_E}$ states so that they cancel, leaving the transition frequency from $^1$S$_0$ insensitive to a large range of tweezer intensities.  

Specifically, by applying a large magnetic field $\vec{B}$ at an angle $\theta$ from the tweezer polarization $\vec{E}_t$, we can induce Zeeman shifts that are much larger than the light shifts associated with the tweezer, so that the eigenstates are primarily defined by the quantization axis $\hat{q}$ oriented along $\vec{B}$ (see Fig. \ref{fig:magic}a).  We label these eigenstates $\ket{-1_B}$, $\ket{0_B}$ and $\ket{1_B}$.  The energy of $\ket{0_B}$ is, to first order, insensitive to the value of the applied field, and can be made insensitive to the light shifts associated with the tweezer.  We can express this state in the basis defined by the tweezer polarization as $\ket{0_B}= \ket{0_E}\cos{\theta} + \frac{1}{\sqrt{2}}(\ket{1_E} - \ket{-1_E})\sin{\theta}$. In the perturbative limit, the light shift due to the tweezer is then $E_0\cos^2\theta + E_1\sin^2\theta$.  If $E_0$ and $E_1$ have opposite signs, then this light shift vanishes at first order for a polarization angle $\tan^2\theta = -E_0/E_1$.  The wavelength at which $E_0$ or $E_1$ vanishes is sometimes referred to as the ``magic wavelength" for such transitions \cite{ido2003recoil}, and represents the edge of the range over which a state-insensitive trapping condition can be reached.  

We apply this technique to generate state-insensitive trapping at tweezer wavelengths near 515~nm.  In Fig.~\ref{fig:magic}b, we measure the angle at which peak excitation from a probe placed on resonance with the non-shifted  $\mathrm{{^1S_0} \leftrightarrow {^3P_1}}$ transition occurs (the ``magic angle") for a set of wavelengths within the tuning range of our laser.  To measure this excitation fraction, we apply the probe light with a Rabi frequency of several tens of kHz for a duration of 100~$\mu$s.  Immediately after the probe light is shut off, we apply a strong ``blow-away'' pulse of light resonant with the  $\mathrm{{^1S_0} \leftrightarrow {^1P_1}}$ transition for 3~$\mu$s, and simultaneously blink off the tweezer light for 1~$\mu$s.  Atoms that were excited to $^3$P$_1$, which has a lifetime of around 20~$\mu$s, do not experience the blow-away pulse and remain trapped, while atoms that were not excited are ejected from the tweezer.  We find that near our tweezer wavelength of 515~nm the magic angle decreases at longer wavelength, indicating a decreasing ratio of $E_0$ to $E_1$ in agreement with calculations based on known levels and transition strengths in strontium \cite{sansonetti2010wavelengths}.  

In Figure~\ref{fig:magic}c, we show the sensitivity of the $^1$S$_0$ to $^3$P$_1$ ($\ket{0_B}$) transition to tweezer intensity by measuring the probe frequency at which peak excitation occurs.  To illustrate the performance and range of applicability of the technique, we do so for three field configurations: a large (24 Gauss) magnetic field oriented at the magic angle for the relevant tweezer wavelength (515.13~nm for this data, corresponding to $\theta = 24(1)$ degrees), a large (22 Gauss) field oriented parallel to the tweezer polarization ($\theta = 0(1)$), and a smaller (7 Gauss) field oriented at the magic angle. In the presence of the non-magic ($\theta = 0(1) $) field, we observe a sensitivity (directly corresponding to $E_0$) of 1.6(1)~MHz/mK.  With the large field at the magic angle, we observe a sensitivity of 8(14)kHz/mK, representing a suppression factor of at least 50 in tweezer-induced shifts compared to the $\theta = 0(1)$ case, and a part in a thousand of the trap depth.  With the smaller field oriented at the magic angle, we observe a low sensitivity to tweezer intensity at low tweezer depths, which increases at higher tweezer depths as the light shifts from the tweezer become comparable to the Zeeman shifts, and the magnetic field no longer defines the character of the eigenstates.  For the relatively deep tweezers that we routinely use, care must be taken to apply a large enough bias field to maintain the perturbative condition required by our technique for state insensitive confinement.

\begin{figure*}[!htb]
\includegraphics[width=\linewidth]{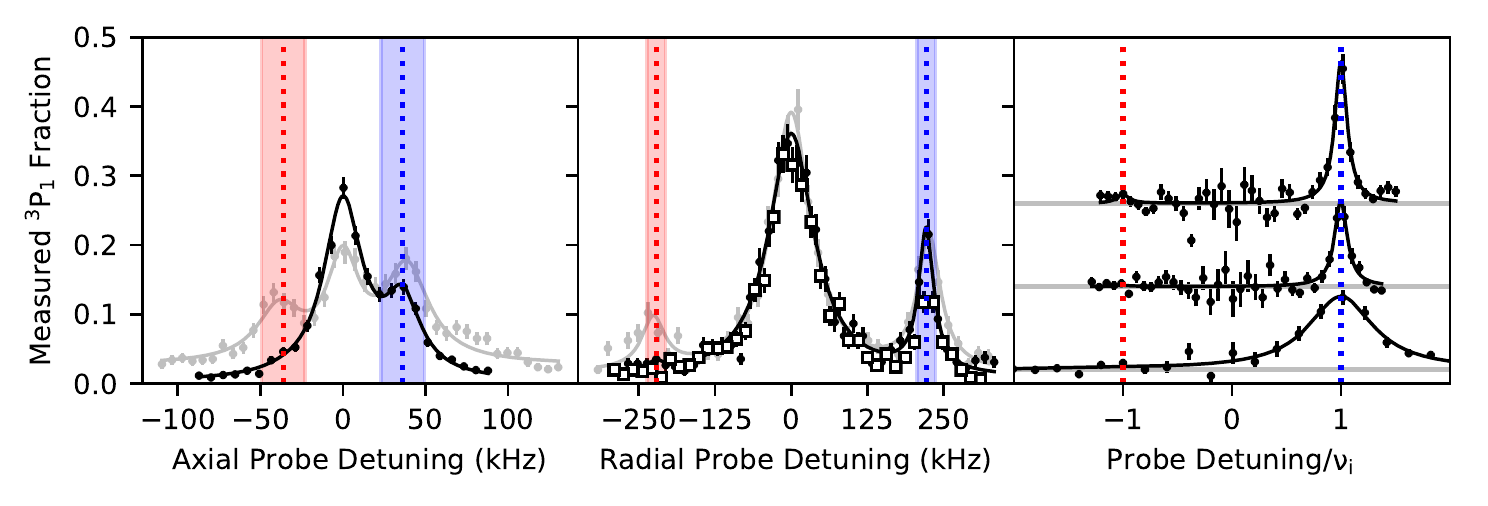}
\caption{Thermometry using sideband spectroscopy.  When performing spectroscopy of the  $\mathrm{{^1S_0} \leftrightarrow {^3P_1}}$ transition after sideband cooling, we observe a clear asymmetry in the strength of the heating and cooling motional sidebands along both the axial (left panel, black trace) and radial (center panel) axes of our tweezer.  These asymmetries correspond to $\bar{n}_{ax} = 0.00^{+0.10}_{-0}$  in the axial direction, and,  $\bar{n}_{r1}=0.06^{+0.26}_{-.06}$ and $\bar{n}_{r2}=0.04^{+0.20}_{-.04}$ in each of the two radial directions probed (corresponding to the black points and line, and white points respectively in the center panel). For comparison, spectra corresponding to trials without sideband cooling (though with narrow-line cooling from the three-dimensional MOT beams) are shown in grey; these indicate $\bar{n}_{ax}<1.3$ and $\bar{n}_{r1,r2}<3$ (see supplement~\cite{Supplement}). The right panel shows the spectra with the fitted contribution from the carrier transition subtracted. From top to bottom, the traces correspond to the first and second radial directions, and the axial direction. These traces are offset from zero for visibility, as indicated by the grey horizontal lines, with the frequency axis scaled by $\nu_i$ (the sideband frequency in the $i$th direction).  Where relevant, the red and blue lines (rectangles) in each panel denote the fitted center (full width half maximum) of the red and blue sidebands respectively.}
\label{fig:sideband_Spect}
\end{figure*}

\section{Resolved-sideband cooling in three-dimensions}

With the magnetic field oriented at the magic angle, we can spectroscopically resolve sideband transitions between different motional eigenstates, which can be labeled in each direction by $n_i$, the number of motional energy quanta in the direction $i$. This allows us to cool the atoms to their motional ground state in all three spatial directions, and to quantify the resulting ground state occupation.  We cool the atoms by driving the $n_i \rightarrow n_i-1$ motional sideband along a given direction ($\Omega_{r1,r2,a}$ defined in Fig.~\ref{fig:setup}a), which removes one motional quantum from that direction.  To cool all three directions, we alternate between the axial and two nearly orthogonal radial directions (see supplement \cite{Supplement}) for a total cooling time of 25~ms.  In what follows, we use a trap depth of 1.2~mK.  

Compared to Raman-sideband cooling of alkali atoms and alkali-like ions \cite{Kaufman2012, thompson2013coherence, monroe1995resolved}, where separate lasers drive the coherent and dissipative steps of the cooling process, the cooling protocol here is relatively simple.
A single laser excites the narrow-linewidth optical transition on the red-sideband, and spontaneous emission returns the atom to its original internal state to enable absorption of further cooling photons~\cite{Blatt2009}. Since the atom is tightly confined in the optical tweezer, the momentum imparted by spontaneous emission is unlikely to cause motional excitation \cite{dicke1953effect}, so that its reduced motional state is also preserved. After several cycles of this process, the atom occupies the ground state, which decouples from the optical drive~\cite{Leibfried2003}. The unique ground state of $^{88}$Sr greatly eases requirements on the polarization purity used in the cooling beams. 

Spectroscopy of the motional sidebands (with the same blow-away procedure used to measure the magic polarization angle) can be used to infer the fraction of atoms that occupy the ground state along a given direction (Fig.~\ref{fig:sideband_Spect}).  Because an atom that is already in the ground state cannot be transferred to a lower motional state, a substantial ground state fraction results in a suppression of the cooling $n_i \rightarrow n_{i-1}$ motional sideband (red sideband, RSB) relative to the heating $n_i \rightarrow n_{i+1}$ sideband (blue sideband, BSB)~\cite{Leibfried2003}; this so-called ``sideband asymmetry" is apparent in all three directions. Assuming a thermal distribution, the average occupation number $\bar{n}_i$ following sideband spectroscopy may be extracted via the expression $A_{RSB}/A_{BSB} = \bar{n}_i/(\bar{n}_i+1)$, where $A_{RSB}$ and $A_{BSB}$ are the fitted heights of the RSB and BSB, respectively (see supplement~\cite{Supplement}). 

Because of the finite excited-state lifetime, photons can be scattered during the sideband spectroscopy probe pulse, potentially modifying $\bar{n}_i$ and the associated ground-state fraction.  This can lead to systematic errors in two main ways: by heating the atoms when probing the BSB, and by cooling the atoms while probing the RSB.

In the first case, the heating increases the measured population transfer on the blue sideband, causing an overestimate of the sideband asymmetry. Indeed, when we double the length of the probe pulse to 200~$\mu$s, $A_{BSB}$ increases by 30\% (in all axes).  We apply a corresponding correction to the BSB height used in our analysis, which increases (reduces) the extracted $\bar{n}_i$ (ground-state fraction).  

The second case of cooling while probing the RSB is an unavoidable possibility. To ease analysis of this process, we use the same laser parameters for the probe pulse as for the cooling that precedes it. Accordingly, the probe pulse on the RSB is a continuation of the cooling along the probed axis. This implies that the RSB height measured at the end of the pulse reflects the total cooling achieved along the probed axis, from both the nominal cooling phase and probe phase. The heating of the un-probed axes from photons scattered during spectroscopy is minimal (1-3\% motional excitation per scattered photon), so that the probing of one axis minimally affects the $\bar{n}_i$ of another~\cite{Supplement}. As such, the combined cooling and probe sequence reflects the final occupation we would achieve had we simply augmented the cooling phase by the duration of the probe phase.

We can further assess to what extent the probe influences the ground-state fraction based on both a master equation calculation and separate analysis of our data. We find that an observed radial $\bar{n}_{r1,r2} = 0.05$ can be at most $0.15$ higher (12\% percent reduction in ground-state fraction) prior to the probe pulse; $\bar{n}_{a}$ is affected by a smaller amount~\cite{Supplement}. While we include this effect in our three-dimensional error analysis, our full set of measurements make it unlikely that the probe pulse significantly affects the ground-state fraction. The $100~\mu s$ probe pulse is short compared to the full cooling duration (25 times longer in each radial direction and 200 times longer in the axial), so that the ground state population measured at the end of the probe pulse is very close to that at the beginning. More precisely, given our initial and final measured ground state fractions (Fig.~\ref{fig:sideband_Spect}), cooling duration, and probe pulse duration, we can bound the change in ground-state fraction that occurs during probing of cooled atoms to below 2\% (see supplement~\cite{Supplement}). Based on these considerations, we conclude that the ground-state fraction directly after cooling is consistently reflected by the probe signal for our experimental conditions, but we include the worst-case possibility reflected by the master equation in our error analysis.

From the data in Figure~\ref{fig:sideband_Spect} and according to the above analysis, we infer a mean excitation number of $\bar{n}_{r1}$ = $0.06^{+0.26}_{-.06}$, $\bar{n}_{r2}$ = $0.04^{+0.20}_{-.04}$ in the two radial directions, and $\bar{n}_{a}=0.00^{+0.10}_{-0}$ in the axial direction. This corresponds to a three-dimensional ground-state fraction of $91^{+9}_{-25}$~\%, which is statistically consistent with master-equations calculations of the ideal performance (96\%)~\cite{Supplement}. We are largely limited in determining the ground-state fraction by the sensitivity and systematics in our thermometry procedure. In the future, more precise determination of the ground state fraction could be made using spectroscopy of the ultra-narrow clock transition (though this would require operation at a magic-wavelength for the clock transition), or from the contrast of multiparticle quantum interference~\cite{Preiss2013,Kaufman2014,Islam2015}.

\section{Single-particle imaging}
\begin{figure}[!htb]
\includegraphics[width=\linewidth, ]{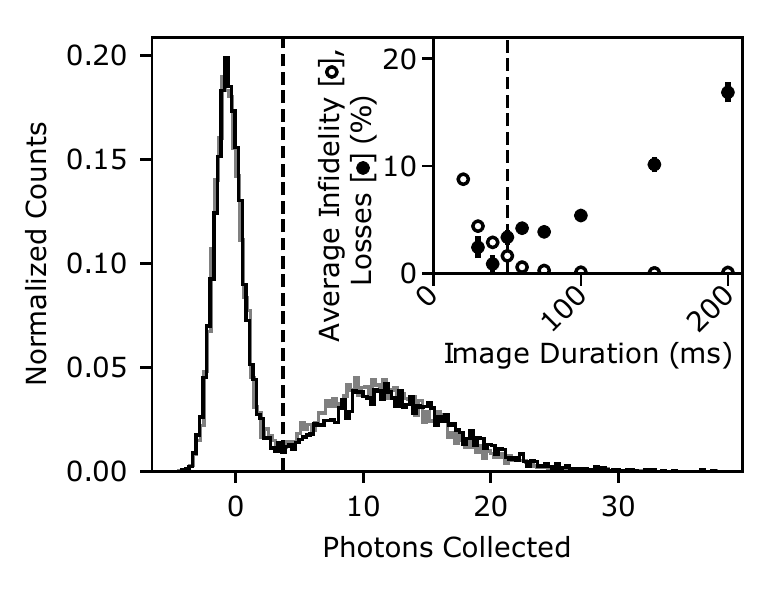}
\caption{Number of photons collected relative to camera background in successive images of a single tweezer (grey trace is first image, black is second). Two peaks are clearly visible, corresponding to zero and one atom in the tweezer.  By applying a threshold (dashed vertical line) at 3.7 photons, we can distinguish an occupied tweezer from an unoccupied tweezer with 98.4(1)\% average fidelity.  We observe 2.4(4)\% atom loss between the two images. Inset: infidelity (empty circles), losses (black points) versus imaging duration. Choosing an imaging duration involves a tradeoff between imaging long enough to collect a sufficient number of photons to resolve the presence of an atom, and the loss associated with scattering many photons. Dashed line denotes where we typically operate in this experiment.
}
\label{fig:hist}
\end{figure}

By utilizing the fast scattering rate and short wavelength of the $\mathrm{{^1S_0} \leftrightarrow {^1P_1}}$ transition at 461~nm, and the cooling properties of the $\mathrm{{^1S_0} \leftrightarrow {^3P_1}}$ transition at 689~nm, we can perform imaging of the atoms with high resolution, high-fidelity and minimal atom loss.  To do this, we continuously apply sideband cooling while pulses of imaging light are applied with a 10\% duty cycle.  The sideband cooling alternates between one radial direction and the axial direction.  

The 461~nm imaging light scattered from the atoms is collected by the objective and imaged onto a cooled, electron-multiplying CCD camera. To determine the presence of an atom in a given tweezer, we define a corresponding region of interest on the CCD, from which we extract the total number of incident photons.  Figure \ref{fig:hist} shows a histogram of counts in such regions over many runs of the experiment.  There is a clear separation between the number of counts observed when an atom is present and when the tweezer is empty, which can be distinguished by defining a count threshold.  Typically, we use a threshold value that minimizes the observed infidelity, as determined numerically.  

Two types of errors can occur during the imaging process -- infidelity errors, where an atom is mistakenly interpreted as an empty tweezer or an empty tweezer is mistaken for an atom, and loss errors, where the atom is lost at some point during the imaging process. We can characterize these processes by examining correlations between two successive images taken in quick succession. We find that for optimized imaging parameters and 50~ms long images, we can distinguish an occupied tweezer from an unoccupied one with 98.4(1)\% average fidelity, and that an atom has a 2.4(4)\% chance of being lost between the two images. These rates represent a balance between collecting enough photons to distinguish the presence of an atom from the camera background and minimizing loss associated with the scattering of imaging and cooling photons. This tradeoff is illustrated in Figure~\ref{fig:hist} (inset), where we vary the duration of the first of a pair of images.  

A significant portion of this atom loss is likely due to decay of the $^1$P$_1$ state into the long-lived and highly anti-trapped $^1$D$_2$ state, which is theoretically predicted to occur roughly once per 20,000 photons scattered on the imaging transition \cite{safronovanote}.  Given our detection efficiency of roughly one in fifteen photons (dominated by the finite numerical aperture of our objective), this mechanism should contribute roughly 1\% loss per 50~ms image.  

We also see evidence of tweezer-light induced loss from the $^3$P$_1$ state used for cooling, which is of comparable magnitude to the expected loss from the $^1$P$_1$ branching when the tweezer depth is around 500~$\mu$K.  We measure that this loss scales approximately linearly with the tweezer depth and with the number of photons scattered on the $^3$P$_1$ transition.  A likely explanation for this loss, which we estimate to be of the correct magnitude, is that the tweezer light causes scattering from the $5s5d$~$^3$D$_{1, 2}$ states which couple to $^3$P$_1$ via a dipole-allowed transition near 487~nm \cite{sansonetti2010wavelengths}.  In principle, these atoms should decay into $^3$P$_1$ (which should then decay to the ground state), or to $^3$P$_{0,2}$, from which they could be recovered by repumping.  However, we have not seen evidence that repumping is effective, perhaps because of the large degree of anti-trapping of the $^3$D$_{1, 2}$ states and large light-shifts of the repumping transitions.  

Given these considerations, we image in relatively shallow (200~$\mu$K) tweezers so as to mitigate the loss mediated by $^3$P$_1$. Importantly, this shallow depth is similar to those used in quantum gas microscopes, suggesting that this technique may be transferable to single-site imaging in lattices~\cite{BlochQGM,Parsons2015}. Furthermore, since the imaging stage in a tweezer experiment can be one of the most demanding of optical power, this shallow depth, corresponding to $\sim 0.4$ mW/tweezer, will aid scaling to large arrays with moderate laser power requirements.  

\section{Conclusion}
We have demonstrated key capabilities for microscopic control of the alkaline-earth atom strontium -- state-insensitive trapping, the ability to perform cooling to the three-dimensional motional ground state using a narrow linewidth transition, and the ability to perform high-fidelity, low-loss imaging of the atoms within the tweezers. These results represent a promising starting point for the implementation of rearrangement techniques for preparation of arbitrary initial states with vanishing entropy~\cite{Barredo2016, Endres2016, Kumar2018}. 

The scalability of our approach for low-entropy lattice systems will depend on the number of traps  achievable with a given tweezer power, and on a more precise determination of our temperature. Incorporating an axial optical lattice will improve the confinement along the axial direction and should improve the cooling robustness and relax the present power requirements of 2 mW/tweezer. With this system in place, we expect to be able to generate several 100 traps based on currently available optical power.

A better understanding of the temperature, and further quantum state control can be gained through spectroscopy of the ultranarrow $\mathrm{{^1S_0} \leftrightarrow {^3P_0}}$ clock transition in tweezers~\cite{Katori2010,Bloom2014}. The long excited state lifetime of this transition will eliminate the effects of cooling and heating during probing, and enable better resolution of the motional sidebands.   These investigations will also provide a first benchmark on coherence for an optical clock transition in a tweezer system.  Ground-state cooled tweezer arrays of strontium with nondestructive imaging could support a new optical atomic clock architecture --- a tweezer array clock ---  that achieves the rapid duty-cycles of ion-based optical atomic clocks, the single-particle isolation of the Fermi degenerate optical lattice clock, along with large particle numbers. It further would be suitable for fundamental studies of squeezing and long-range entanglement on an optical clock transition~\cite{Rosenband2008, Pohl2014, Zeiher2017, campbell2017fermi}.

\section{Acknowledgements}
We acknowledge fruitful discussions with  M. Greiner, D. Greif, M. D. Lukin, J. K. Thompson, C. A. Regal, A. M Rey, and J. Ye. We would like to thank B. Johnston and W. J. Eckner for technical assistance. We further acknowledge helpful collaborative communication with the group of M. Endres. Lastly, we thank R. B. Hutson of the Ye group for sharing his polarizability calculations.  This work was supported by the National Science Foundation Physics Frontier Center at JILA (1734006) and the National Institute of Standards and Technology. 

\clearpage

\bibliography{references.bib}

\pagebreak
\begin{center}
\textbf{\large Supplemental Materials}
\end{center}
\setcounter{section}{0}
\setcounter{equation}{0}
\setcounter{figure}{0}
\setcounter{table}{0}
\setcounter{page}{1}
\makeatletter
\renewcommand{\theequation}{S\arabic{equation}}
\renewcommand{\thefigure}{S\arabic{figure}}

\section{Summary of Apparatus}

\begin{figure}[!htb]
\includegraphics[width=\linewidth, ]{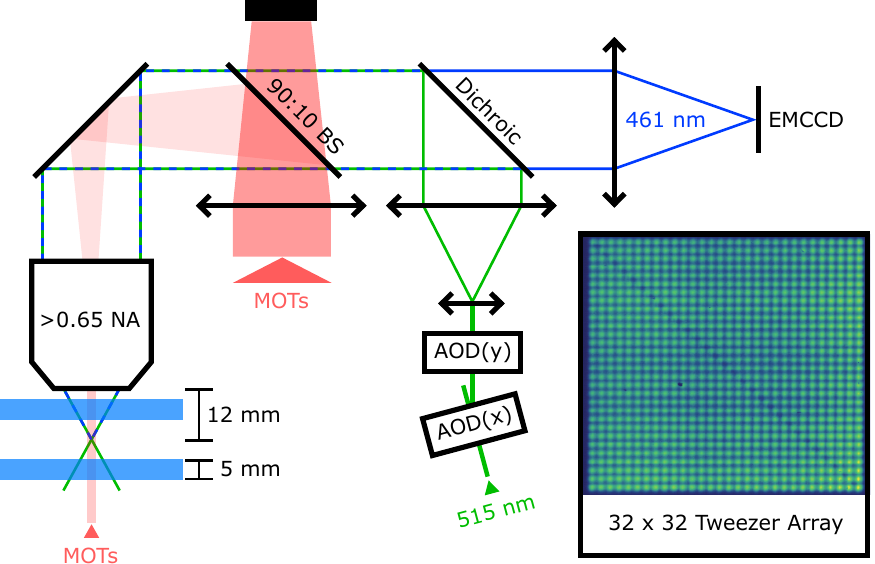}
\caption{Simplified schematic of optics layout. The vertical MOT beams (at both 689~nm and 461~nm) are combined onto the beam path with a custom 90:10 beam splitter. The 515~nm tweezer light is combined onto this path with a dichroic designed to be maximally transmissive at 461~nm and reflective at 515~nm. Inset: preliminary example of 32$\times$32 tweezer array.}
\label{fig:layout}
\end{figure}

The optical layout of our experiment is summarized in Fig.~\ref{fig:layout}. A high numerical aperture (NA $>0.65$) objective lens with a 12~mm working distance from the atoms is used both to project tightly confining optical potentials at $\sim$515~nm to load strontium atoms into, and to image the atoms on the $\mathrm{{^1S_0} \leftrightarrow {^1P_1}}$ transition at 461~nm. A pair of crossed acousto-optic deflectors (AODs) are imaged onto each other in a standard 4f configuration to generate deflections in two orthogonal directions, forming an array of tweezer spots in the image plane of the objective.

This high-NA system makes it challenging to address the atoms with beams that have an appreciable wave vector component along the axis of the objective. This is particularly important in the case of the vertically confining magneto-optical trap (MOT) beams which, in addition to having a projection along this axis, must also have a large mode-field diameter. To address this, we focus our vertical MOT beams at the back-focal plane of the objective, so that they exit the objective collimated. These are paired with collimated beams launched from below the objective to form the vertical axis of our 3D MOTs.

The performance of our imaging system can be characterized by fits to the point spread function (PSF) of our atom images. These yield an effective Gaussian waist of 0.44(2)~$\mu$m at 461~nm as shown in Fig.~\ref{fig:setup}c of the main text. This measurement should be treated as an upper bound, as it is sensitive to vibrations, air currents, and long-term drifts in the imaging system.

To estimate the waist of our 515~nm tweezers, we compare the radial trap frequencies measured during spectroscopy to the expected trap frequencies based on the known power per tweezer spot and polarizability of the $\mathrm{^1S_0}$ state of strontium. Near 515~nm, we use a polarizability of $\mathrm{900}~a_0^3$ in atomic units (this can be converted to SI units by the conversion factor $4\pi\epsilon_0 a_0^3/h$), where $a_0$ is the Bohr radius~\cite{Safronova2015}. This analysis yields an effective Gaussian waist of 0.48(2)~$\mu$m.

\section{Experimental Sequence}

Here we summarize the standard sequence of operations in the experiment, including initial trapping, state preparation, and atom detection.

\begin{figure*}[!htb]
\includegraphics[width=\linewidth, ]{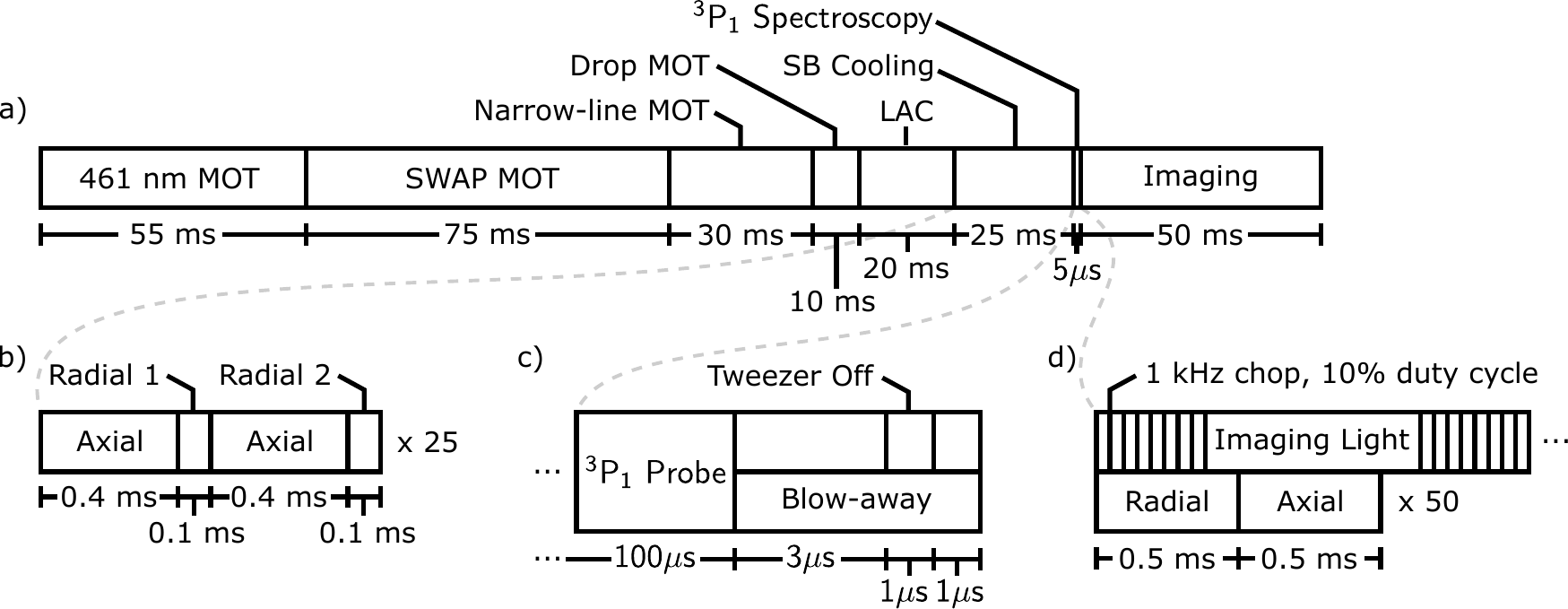}
\caption{Characteristic timing diagram. (a) Full experimental sequence for taking $\mathrm{^3P_1}$ spectra. (b, c, d) detail views of sideband cooling, $\mathrm{^3P_1}$ transfer, and imaging sequences respectively. }
\label{fig:sequence}
\end{figure*}
\subsection{Initial cooling and loading}

The full experimental sequence is summarized in Fig.~\ref{fig:sequence}. Our experiments begin by capturing a thermal beam of $\mathrm{^{88}Sr}$ atoms in a MOT operating on the 32~MHz wide $\mathrm{{^1S_0} \leftrightarrow {^1P_1}}$ transition at 461~nm, which cools the atoms to a Doppler-limited temperature of $\mathrm{\sim1}$~mK. We release the atoms from the 461~nm MOT, and recapture them in a narrow-line MOT operating on the $\mathrm{{^1S_0} \leftrightarrow {^3P_1}}$ transition at 689~nm that is formed using sawtooth-wave adiabatic passage (SWAP MOT)~\cite{norcia2018narrow,muniz2018robust}.  In this stage, the three-dimensional cooling beams, which co-propagate along the 461~nm MOT beams, are swept upwards in frequency over a range of 6~MHz with a repetition rate of 30~kHz, ending at a frequency of +100~kHz relative to the zero-field $^1$S$_0$ to $^3$P$_1$ transition.  We have measured the temperature during this stage to be roughly 35~$\mu$K. The tweezers are on during the whole MOT portion of the experimental sequence.  To load from the SWAP MOT into the tweezers, we simply ensure that the atoms are spatially overlapped with the tweezers, and leave the SWAP MOT on for 75~ms. For most experiments, we perform an additional 30ms of cooling with an unswept narrow-line MOT. We then drop the narrow-line MOT to isolate the atoms that have been trapped in the tweezers from the background gas.

\subsection{Light-assisted collisions}

The atom densities and temperatures achieved in our narrow-line MOT often result in multiple atoms occupying each tweezer after loading. For our typical experiments, we ensure that we begin with either one or zero atoms in each tweezer by inserting an additional step after loading with identical parameters to our imaging sequences (pulsed 461~nm light combined with sideband cooling), though with a shorter duration of 20~ms. Light assisted collisions that occur during this step cause pairwise loss, resulting in either zero or one atom in each tweezer~\cite{Schlosser2001}. With this step in place, we typically observe a fill fraction near 50\% (45-50\% typical), and do not see evidence of multiply loaded sites in count histograms obtained during imaging.

\subsection{Ground-state cooling protocol}

In order to cool single atoms to the three-dimensional mechanical ground state of the tweezers, we alternate between addressing the cooling sidebands in the axial direction, and two nearly orthogonal radial directions, as shown in Fig.~\ref{fig:sequence}b. We repeat this cycle 25 times before our spectroscopy measurements. The intensities used in each direction are equal to those used for the spectroscopy data shown in Fig.~\ref{fig:sideband_Spect} of the main text, corresponding to carrier Rabi frequencies of 60~kHz in the radial directions and 10~kHz in the axial direction.  We have measured the loss rate associated with this cooling sequence to be 1(1)\% when applied between two imaging sequences.  

\subsection{Spectroscopy protocol}
In order to perform spectroscopy of the $\mathrm{{^1S_0} \leftrightarrow {^3P_1}}$ transition, we apply a pulse of light with tunable frequency along a given direction (the probe pulse).  Typically, we scan this frequency over the two first-order motional sidebands and the carrier over a series of experimental trials.  We use the same Rabi frequencies as those described in the cooling section for each axis, and a pulse duration of 100~$\mu$s in order to resolve spectrally narrow features.  

In order to determine if an atom was transferred to $\mathrm{^3P_1}$, we apply a ``blow-away" pulse of light near resonance with the 461~nm $\mathrm{{^1S_0} \leftrightarrow {^1P_1}}$ transition.  The total duration of this pulse is 5~$\mu$s.  During this pulse, we also blink off the tweezer for 1~$\mu$s, as summarized in Fig.~\ref{fig:sequence}c.  We find this effective at ensuring that ground state atoms are removed with a probability of at least 98\%.  Because the duration of the blow-away pulse is significant compared to the $\mathrm{^3P_1}$ lifetime of 20~$\mu$s, we also lose some portion of the atoms that were transferred to $\mathrm{^3P_1}$. The highest fraction of atoms remaining after the blow-away pulse is 40\%, compared to the expected 50\% peak transfer fraction, so we estimate that 80\% of atoms that were excited to $\mathrm{^3P_1}$ survive the blow-away pulse.  
Because we are interested in ratios of excitation probabilities in thermometry, this loss does not impact our results. 

\subsection{Imaging}
During imaging, we continuously apply sideband cooling, while applying pulses of light near resonance with the $^1$S$_0$ to $^1$P$_1$ transition at 461~nm.  The sideband cooling alternates between the axial direction and a single radial direction, with each applied for 0.5~ms at a time (Fig.~\ref{fig:sequence}d). For the axial cooling, we find it most effective to red detune by roughly 100~kHz from the axial carrier transition and drive higher-order cooling sidebands (which are not clearly resolved from one-another for the intensities used). The 461~nm light is pulsed with a duty cycle of 10\%, at a frequency of 1~kHz.  In order to reduce potential heating from the anti-trapped $^1$P$_1$ state, we detune the imaging light by roughly 600~MHz from the free-space transition frequency \cite{cheuk2015quantum}.  

As described in the main text, we detect the presence of an atom by collecting the 461~nm photons on a cooled, electron-multiplying CCD camera (EMCCD, model: Andor iXon 897).  We integrate the total number of photons collected from each atom over the corresponding region of the sensor, and define a threshold count number above which we infer the presence of an atom. We characterize the fidelity and loss rate of our imaging process by taking two images in quick succession, and analyzing the correlations between the images. 

Formally, we define $P(0)$ ($P(1)$) as the probabilities that an atom is absent (present) in the first image on a given trial.  $P(0|1)$ and $P(1|0)$ are the probabilities that a void is mistaken for an atom and that an atom is mistaken for a void, respectively.  $P(1|1)$ and $P(0|0)$ correspond to the probabilities of correctly identified atoms and voids.  $P_{\rm{loss}}$ is the probability that an atom that was present in the first image is lost between the two images (we assume that an atom cannot appear between the two images), and $P(XY)$ represents the probability of measuring $X$ in the first measurement and $Y$ in the second, where $X$, $Y$ can be either 0 or 1, indicating a void or an atom.  We assume that $P(0|1)$ and $P(1|0)$ are small, and neglect terms that involve products of these quantities from future expressions.  We also assume that $P(1|1)$ and $P(0|0)$ are near unity.  These assumptions allow us to obtain simple expressions for the average infidelity and loss, and contribute errors that are small relative to the statistical uncertainty on our measurements.

We can infer the average infidelity from the fraction of the trials in which we measure a void in the first image and an atom in the second $P(01)$.  This accounts for both trials in which an atom was present in both images, but was mistaken for a void in the first image ($P(0|1)$), and for events where no atom was present in either image, but a false count was recorded in the second ($P(1|0)$):

\begin{equation}
\begin{split}
\begin{aligned}
&P(01) \simeq  P(1)P(0|1)P(1|1) + P(0)P(0|0)P(1|0) \simeq \\ &P(1)P(0|1)+P(0)P(1|0)
\end{aligned}
\end{split}
\end{equation}
\noindent We define this quantity as the average infidelity, representing the probability of making an error averaged over initial conditions.  

The loss can be inferred from the excess number of trials where a void follows an atom relative to the number of trials in which an atom follows a void:

\begin{equation}
\begin{split}
\begin{aligned}
&P(10) \simeq \\ &P(0)P(1|0)P(0|0) + P(1)P(1|1)P(0|1) + \\ &P_{\rm{loss}}P(1|1)P(0|0) \simeq \\ &P(01) + P(1)P_{\rm{loss}}
\end{aligned}
\end{split}
\end{equation}

\section{Theoretical analysis of spectroscopy and ground-state cooling}
\label{supp:master}

\begin{figure}[!htb]
    \includegraphics[width=\linewidth]{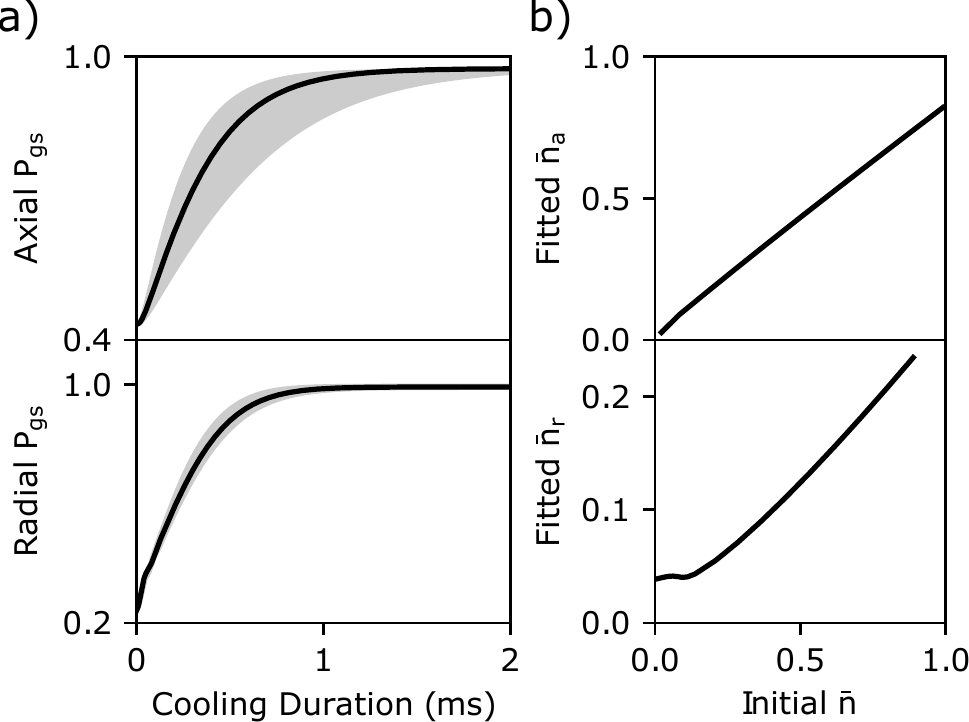}
    \caption{Master equation calculations. (a) Ground state population ($P_{gs}$) trajectory during cooling starting from upper bounds for the atom temperature after loading in the experiment. Simulations carried out in the axial (radial) direction for Rabi frequencies of 8.5~kHz (65~kHz) with a confidence interval of $\pm 2.5$~kHz ($\pm 10$~kHz) denoted by grey region. (b) We simulate the spectra for the axial and radial probing parameters for varying initial $\bar{n}$ ($x$-axis). We then fit the resulting simulated spectra identical to how we fit our data in order to extract the fitted $\bar{n}$ ($y$-axis). At low initial $\bar{n}_{r1,r2}$, no further cooling is achieved given the probe parameters (i.e. the initial value is at or below the equilibrium value associated with the probe frequency being at the RSB frequency).}
    \label{fig:suppCooling}
\end{figure}

 We perform numerics with a one-dimensional master equation to model the combined coherent and dissipative dynamics associated with the spin and motional degrees-of-freedom when driving the ${^1S_0} \leftrightarrow {^3P_1}$ transition~\cite{Leibfried2003, Kaufman2012, Daley2014}. The calculation takes as input the  Rabi frequencies for the red cooling beam, trap frequencies (40~kHz and 220~kHz for the axial and radial axes, respectively), and the excited state decay rate $\Gamma = 2\pi\cdot 7.5~\mathrm{kHz}$; we truncate the Hilbert space at 8-10 vibrational levels and use a spin-1/2 system.

\textbf{Ground-state fraction from the master-equation} In Figure~\ref{fig:suppCooling}a, we show simulated cooling trajectories for the radial and axial dimensions of the trap. These predict that $> 97$\% ($> 99$\%) ground state fractions can be obtained in the axial (radial) directions. We include an uncertainty band to illustrate that these results are robust to changes in the Rabi frequency; for the radial direction, they reflect the range used for each radial axis. For comparison, the experimental timescale for the total cooling applied along each axis is 10 (100) times longer than the 1/e-timescales indicated for equilibrium. We note that the master-equation time-scales represent the ideal case; we expect that the cooling rate is slower than reflected by these plots due to the effects of laser-noise, which are not included in the numerics. 

\textbf{Cross-coupling during probing} The master equation calculation represents an upper-bound on the ground-state fraction, because it does not include the effect of cross-coupling between the axes, that is, the degree of heating along one axis when cooling along another. This heating can come about in two ways: recoil heating of one axis while another is cooled, and latent heating in the traps. We quantify these effects in order to understand how well the one-dimensional master equation describes our three-dimensional system. 

When cooling, in order to remove one motional quanta from an axis, a photon must be scattered into free space. This photon
can heat another axis. Assuming isotropic photon emission, the probability that this photon heats a given axis is $\eta_i^2/3$ where $\eta_i$ is the Lamb-Dicke parameter in the $ith$ direction. In our system, $\eta_a = 0.35$ ($\eta_{r1,r2} = 0.15$) in the axial (radial) direction. Accordingly, a 689~nm photon scattered by the atom has a $3\%$ ($0.8\%$) chance of heating by one motional quantum along the axial (either radial) axis.  When the atom is already cold, the rate of photons scattering from the RSB is also suppressed.  Based on our measured sideband heights, the probability of heating one axis due to a cooling cycle or probe pulse applied to another axis is bounded at the 1\% level. 

Secondly, when cooling one axis, any residual heating rate along another axis in the trap will remain uncompensated. We can measure these heating rates: we observe $<1$ phonon/20 ms in the radial direction, and no observable heating on a 100 ms scale for the axial direction. The heating rate in the radial direction corresponds to a 0.03 increase in a radial axis $\bar{n}_{r1,r2}$ when other axes are cooled.

\textbf{Theoretical limit on temperature change during spectroscopy} We can simulate the effect of the spectroscopy on the temperature by fitting simulated spectra and comparing them to the initial temperature. In Fig.~\ref{fig:suppCooling}b, we plot on the $y$-axis the fitted $\bar{n}_i$ from a simulated spectrum using the master equation, against the initial $\bar{n}_i$ on the $x$-axis. The change can be a significant effect, particularly on the radial direction. Assuming a fitted $\bar{n}_{r1,r2} = 0.05$ in the spectroscopy, these simulations imply the $\bar{n}_{r1,r2}$ prior to spectroscopy could be $0.2$, corresponding to a $12$\% reduction in the ground-state fraction compared to the ground-state fraction implied by the sideband spectroscopy. We fit the linear regimes of these simulations to extract scale factors so that our error bars reflect this phenomenon. However, note that we describe below in Sec.~\ref{supp:dataExp} that we think such a change over the course of spectroscopy is unlikely given our full set of measurements, meaning that the occupation before spectroscopy is very nearly the same as that after.

\section{Experimental analysis of sideband thermometry}
\label{spect}
 
\textbf{Fitting of spectra} We extract the sideband ratio by fitting the sum of three Lorentzians to the sideband spectroscopy data, one corresponding to the carrier transition, and the other two corresponding to the red and blue sidebands (RSB and BSB respectively); these fits provide the amplitudes of the two sidebands, $A_{BSB}$ and $A_{RSB}$.  Because the RSB is typically not resolvable from the background counts for cooled atoms, we impose the condition that the spacing between each sideband and the carrier is equal, and that the two sidebands have equal width. We infer the $\bar{n}$ at the end of the pulse by using the relation that the ratio between the population transferred on the RSB to the population transferred on the BSB is given by $A_{RSB}/A_{BSB}=\bar{n}/(1+\bar{n})$~\cite{Leibfried2003}. 

\label{supp:analysisSpec}

As noted in the text, and analyzed theoretically in Sec~\ref{supp:master}, because the duration of the probe pulse is longer than the excited state lifetime, it is possible that the probe itself either heats or cools the atoms when probing the BSB or RSB, respectively. We discuss below how this phenomenon informs our analysis for the cases of before and after sideband cooling separately. 

\textbf{Analysis of spectroscopy before sideband cooling} In Figure~\ref{fig:sideband_Spect}, we show sideband spectroscopy at the stage of the experiment right before sideband cooling (light gray). For these initial temperatures, the spectroscopy probe pulse can significantly influence the $\bar{n}_i$. For the hottest case given our uncertainties, the temperatures from these data correspond to less than $\bar{n}_{a}=1.3$ ($\bar{n}_{r1,r2}=3$) in the axial (radial) directions according to the master equation model of cooling and heating achieved during the probing. We use these as bounds on our initial temperatures.

\textbf{Analysis of spectroscopy after sideband cooling}
After sideband cooling, we observe the three spectra displayed in Figure~\ref{fig:sideband_Spect} of the main text. The master equation results of Fig.~\ref{fig:suppCooling}b in the linear regime suggest the $\bar{n}_i$ after cooling, but prior to probing, can be a factor of 3.76 (1.23) higher than the fitted $\bar{n}_i$ extracted after probing. However, as we describe below in Sec.~\ref{supp:dataExp}, given the cooling duration, probe duration, and final observed sideband assymmetries, these factors would suggest $\bar{n}_{r1,r2}$ and $\bar{n}_{a}$ \emph{before sideband cooling} that are inconsistent with the spectroscopy measurements before sideband cooling. Given this full set of information, we incorporate the master equation result so that our error bars reflect the hottest possibility. For quoting the center values of $\bar{n}_{r1,r2,a}$,  we assume that the RSB height is unchanged during spectroscopy, while we correct for the observed $30$\% change in the BSB height as discussed in the main text.
 

\textbf{Influence of laser frequency noise}
The finite linewidth of our 689~nm laser results in a reduced scattering rate from the sidebands during both cooling and sideband thermometry.  This reduced scattering rate is evident from the reduced sideband excitation fraction measured in our thermometry scans.  We estimate the linewidth of the laser to be 10-20~kHz based on the observed widths of spectroscopic features.  This range includes the effects of both short-term variations in the laser frequency and drifts over the course of a data-set. The primary effect of this laser noise on the cooling is to increase the length of time it takes to reach equilibrium, which motivates our long cooling duration compared to the time-scales reflected by Fig.~\ref{fig:suppCooling}. For the spectroscopy, it reduces the transfer fraction on the sidebands. 

\subsection{Data-based expectation of probe's effect on temperature} 
\label{supp:dataExp}
Using our spectroscopy data, we can form an expectation for the degree to which the probe influences the perceived height of the RSB using the spectroscopy before and after sideband cooling. The initial temperature is less than $\bar{n}_{r1,r2} < 3$ in each radial direction and $\bar{n}_{ax} < 1.3$ in the axial, while at the end of probing we observe $\bar{n}_{r1,r2} <.2$ and $\bar{n}_{ax} < .1$ (excluding the master equation increase of the error bars).  From a rate equation picture, we would expect the ground-state fraction follows an exponential during the combined cooling and probe sequence. That is, we assume the ground-state fraction along an axis $i$ varies during interrogation on the RSB as $P^i_{gs}(t) = \Delta {P^i_{gs}(1-e^{-t/\tau^i}) + P^i_{gs,in.}}$, where $\Delta P^i_{gs}$ is the change in ground-state fraction in the long-time limit, $P^i_{gs,in.}$ is the initial ground-state fraction, and $\tau^i$ the exponential time-scale. We can constrain $\Delta P^i_{gs}$ and $P^i_{gs,in.}$ based on our measurements of initial and final ground-state occupation from the spectroscopy of each axis. Accordingly, cooling that occurs during probing is less than 2\% of a quantum along the radial direction for all possible $\tau^i$, and much less than 1\% of a quantum along the axial direction. The master equation calculation in Fig.~\ref{fig:suppCooling} is consistent with this conclusion of minimal perturbation (i.e. the change in the ground-state fraction over the last $100\mu $s of cooling is very small), but the in-situ thermometry includes all possible effects that reduce our cooling rate below the optimal case, such as laser frequency noise. We note that this one-dimensional analysis relies on negligible cross-coupling, as justified in Sec.~\ref{supp:master}.

\section{Modeling of the transition shifts}

To model the dependence of the  transition shift of the $\mathrm{{^1S_0} \leftrightarrow {^3P_1}}$, $m=0$ transition, we must determine the eigenvalues of the $^3P_1$ manifold. These are determined by the full Hamiltonian $H$ for the excited state,
\begin{equation}
    H = H_{\vec B} + H_{\vec E},
\end{equation}
where  we have broken up the expression component-wise into the magnetic-field part $H_{\vec B}$ and electric-field part $H_{\vec E}$. The magnetic component can be written in the standard fashion,
\begin{equation}
    H_{\vec B} = \mu_B g_J \vec J \cdot \vec B, 
\end{equation}
where $\vec J$ is the sum of the electron spin and orbital angular momentum, $\vec B$ is the applied magnetic field, $\mu_B$ is the Bohr-magneton, and the gyromagnetic ratio $g_J = 3/2$. Meanwhile, the electric component can be written as~\cite{LeKien2013}, 
\begin{equation}
    \begin{split}
            H_{\vec E} = &-\frac{1}{4}\abs{E}^2(\alpha^s_{e}+\alpha^v_{e}\frac{(\vec \epsilon^* \cross \vec \epsilon) \cdot \vec J}{2J}\\
    &+\alpha^t_{e}\frac{3((\vec \epsilon^*\cdot \vec J)(\vec \epsilon\cdot \vec J)+ (\vec \epsilon^*\cdot \vec J)(\vec \epsilon\cdot \vec J)-2 J^2)}{2J(2J-1)}), 
    \end{split}
\end{equation}
where $\vec \epsilon$ is the tweezer polarization in the basis of spherical components, $\abs{E} ^2$ is the norm-squared of the electric-field, and $\alpha^s_{e}$, $\alpha^v_{e}$, $\alpha^t_{e}$ are the scalar, vector, and tensor polarizabilities, respectively. 

\begin{figure}[!htb]
\includegraphics[width=\linewidth, ]{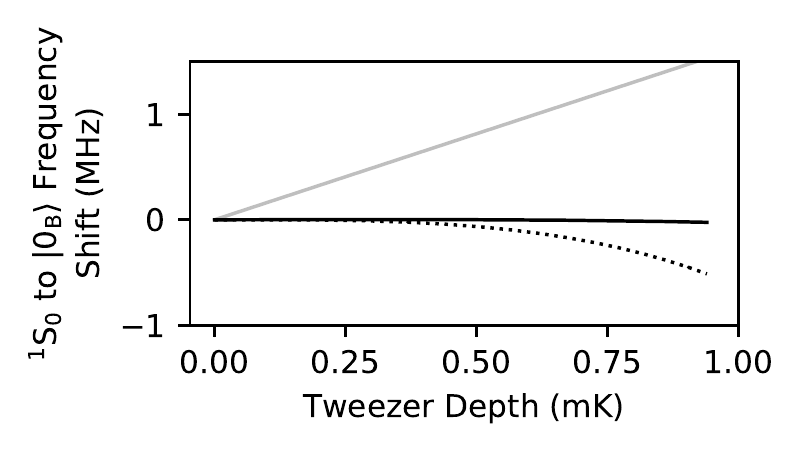}
\caption{We plot the frequency shift on the $^{1}S_0 \leftrightarrow ^3P_1$, m=0 (or $\ket{0_B}$) transition for the experimental conditions described in the main text: $\theta = 0$ (gray), $\theta = \theta_{magic}$ with a 24 G (black) and 7 G (dashed) magnetic field. These plots use $\alpha^S = 760~a_0^3$, $\alpha^T = 140~a_0^3$, and the ground-state polarizability quoted in the text.}
\label{fig:shifts}
\end{figure}

To confirm the perturbative description in the text, we perform exact diagonalization of $H$ (Fig.~\ref{fig:shifts}). From this model, we can extract approximate values of $\alpha^s_{e}$ and $\alpha^t_{e}$ based on the data shown in Fig.~\ref{fig:magic}c of the main text, assuming zero ellipticity in the trap, exact knowledge of our tweezer waist, and with knowledge of the ground-state polarizability. We note that approximate polarizability values are consistent with an ab-initio calculation of the polarizabilities~\cite{Zhou2010} at the $25\%$ level when using a ground-state polarizability also extracted from this calculation. Importantly, Figure~\ref{fig:shifts} confirms our interpretation of the data and perturbative description: there is linear sensitivity at zero angle between the tweezer and magnetic field, there is magic-behavior over a range of tweezer depths at the magic angle, and at reduced magnetic fields the tweezer depth can lead to non-perturbative shifts and breaking of the magic-angle suppression. The key advantage of this technique is that it only depends on the relative sign and magnitude of the scalar and tensor shift as opposed to their exact values, and can be achieved by appropriate application of a magnetic field.

\end{document}